# An Atom Michelson Interferometer on a Chip Using a Bose-Einstein Condensate

Ying-Ju Wang,[1] Dana Z. Anderson,[1] Victor M. Bright,[2] Eric A. Cornell,[1] Quentin Diot,[1] Tetsuo Kishimoto,[1*]
Mara Prentiss,[3] R. A. Saravanan,[2] Stephen R. Segal,[1] and Saijun Wu[3]

[1] *Department of Physics, University of Colorado, and JILA, National Institute of Standards and Technology and University of Colorado, Boulder, Colorado 80309-0440, USA*
[2] *Department of Mechanical Engineering, University of Colorado, Boulder, Colorado 80309-0427, USA*
[3] *Department of Physics, Harvard University, Cambridge, Massachusetts 02138, USA*

An atom Michelson interferometer is implemented on an "atom chip." The chip uses lithographically patterned conductors and external magnetic fields to produce and guide a Bose-Einstein condensate. Splitting, reflecting, and recombining of condensate atoms are achieved by a standing-wave light field having a wave vector aligned along the atom waveguide. A differential phase shift between the two arms of the interferometer is introduced by either a magnetic-field gradient or with an initial condensate velocity. Interference contrast is still observable at 20% with atom propagation time of 10 ms.

Atom interferometry offers exquisite precision measurement capability, much like its photon-based counterpart. For example, the fundamental limit on the atom Sagnac gyroscope signal-to-noise ratio is a factor of $10^{11}$ greater than its optical counterpart, given comparable enclosed areas and particle flux. Unlike their photon counterparts, however, atoms can be sensitive to electric and magnetic fields. In this respect, atom interferometry may be suited to a substantially larger set of sensor applications, though by the same token, it is sensitive to a larger variety of detrimental noise sources. Atom interferometry has been demonstrated using both normal atoms and Bose-Einstein condensates. On balance, atom interferometry experiments have revealed promising and sometimes stunning measurement capabilities [1–5].

Efforts to incorporate interferometry on an "atom chip" [6–9] are motivated by the large physical size of a traditional apparatus and a desire to better tailor interferometer geometries. Most attempts to implement a coherent beamsplitter/recombiner on a chip have used current-induced magnetic fields, typically forming double potential wells that merge and then split apart either in space, in time, or in both. Various technical issues, such as noise coupled into the current and roughness or impurities of the wires, have stymied attempts to demonstrate on-chip interference [10–13].

This work reports the demonstration of an on-chip atom Michelson interferometer employing a Bose-Einstein condensate (BEC). An intrawaveguide optical standing wave serves to split, reflect, and recombine the BEC. We introduce a differential phase shift between the two arms of the interferometer that modulates the atom interference. This differential phase shift is introduced by a magnetic field gradient and alternatively by an initial condensate velocity in a trap with a longitudinal frequency of 5 Hz. We observe interference when the round-trip propagation time is relatively short, i.e., less than about 10 ms and the maximum separation of the split wave packets is about 120 $\mu$m. The maximum separation is greater than the results reported previously for confined atom interferometers [14, 15].

Our atom chip contains lithographically patterned wires and a pair of prism-shaped mirrors, as shown in Fig. 1. The wires generate magnetic fields that are essential for the microtrap and magnetic waveguide. The inward-facing surfaces of the two prisms are mirror-coated and aligned to produce optical standing waves. A 180-$\mu$m-high tunnel is created underneath the prism located at the entrance side of the chip to allow for loading of the microtrap.

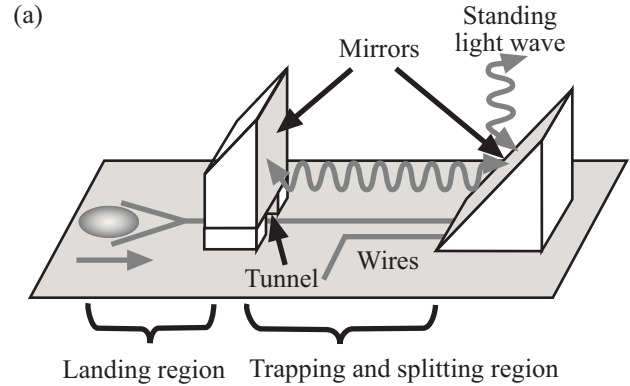

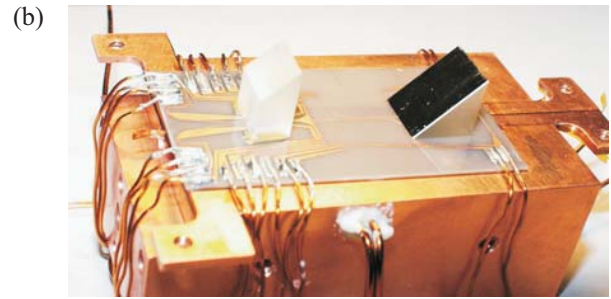

FIG. 1. (a) Schematic drawing of the atom chip (not to scale). The prism-shaped mirrors are integrated with microfabricated wires on an aluminum nitride substrate. The dimensions of the whole chip are 5 cm by 2 cm. (b) A photo image of the atom chip glued onto a copper holder.

A cold cloud of $^{87}Rb$ in the $\left| F=1, m_f=-1 \right\rangle$ state is collected and precooled in adjoining vacuum chambers to around 0.45 $\mu$K [16]. The cloud is delivered through the tunnel and captured in the chip's trapping region, which is located between the prisms and sits 115 $\mu$m away from the surface of the chip. After the cloud is loaded onto the chip, the trapping currents are reduced to place the cloud to 250 $\mu$m away from the surface. Next, rf-forced evaporative cooling is used to form a condensate. The frequencies of the magnetic trap in which the condensate is made are 177 Hz in the radial and 10 Hz in the axial direction. Once a condensate is formed, the axial confinement is ramped down in 100 ms to release the condensate to the waveguide.

A tightly focused, linearly polarized laser beam with a waist of 110 $\mu$m is incident on and reflected by the mirror that is oriented at 45 degrees with respect to the chip surface [see Fig. 1]. The reflected beam is perpendicular to the surface of the chip and is parallel to the direction of the waveguide. The beam is retroreflected by the second mirror that is perpendicular to the waveguide and the surface of the chip. The wave vector of the standing-wave light field needs to be parallel to the waveguide to minimize radial excitations of the BEC from photon scatterings. The prisms must be aligned with respect to the waveguide to better than 2 degrees.

An interferometric measurement involves three standing-wave light pulses: a splitting pulse, a reflection pulse, and a recombining pulse. The frequency of the standing-wave light field is 7.8 GHz red detuned from the atomic resonance to minimize spontaneous emission. The condensate cloud begins at rest in the guide. The splitting pulse actually consists of a pair of subpulses, and each single subpulse diffracts atoms like an optical diffraction grating in the Raman-Nath regime. As a result the cloud is diffracted into different momentum states. The momentum of the atoms diffracted into the nth order is changed by $\pm 2n\hbar k$, where $n$ is an integer number and $k$ is the wave number of the photons [17].

In general, a single light pulse excites several different diffraction orders, which implies a coupling efficiency less than 100% into a particular order. To achieve a nearly ideal 50/50 splitting ratio, we have developed the double subpulse scheme [18]. The condensate starts at rest. Then the first pulse couples some of atoms into the $|p=\pm 2\hbar k\rangle$ state while the others stay in the $|p=0\rangle$ state. The phases of atoms with different momenta evolve at different rates before the second pulse mixes them again. The phase evolution of atoms in the $|p=\pm 2\hbar k\rangle$ state is faster than that of the ones in the $|p=0\rangle$ state. The differential phase shift between the atoms can be written as $\Delta\phi = 4\omega_r \Delta t$, where $\omega_r = \hbar k^2/2m$ is the recoil frequency, $m$ is the mass of atoms, and the $\Delta t$ is the delay between the raising edges of the two pulses. Atoms in the different momentum states interfere as the second subpulse mixes them again. If the phase difference $\Delta\phi$ is a multiple of $2\pi$, which corresponds to the delay $\Delta t = n\pi/2\omega_r$, between the two pulses, most atoms will populate in the $|p=\pm 2\hbar k\rangle$ state. On the other hand, if the phase difference is $(2n+1)\pi$, which corresponds to the delay $\Delta t = (2n+1)\pi/4\omega_r$, most atoms will remain in the zero momentum state after the double pulse. The contrast ratio of this interferometric splitting can be made 100% by optimizing the power and the length of the pulses [18]. In our experiment, splitting of the atoms into the $|p=\pm 2\hbar k\rangle$ states, with nearly zero population in the $|p=0\rangle$ or $|p=\pm 4\hbar k\rangle$ state, can be achieved; we find optimum splitting when the two pulses are both 20 $\mu$s in duration with power around 5.5 $\mu$W and the delay between the two pulses is 63 $\mu$s. The atoms in the different momentum states are detected by absorption imaging when the wave packets are spatially separated after 10 ms propagation.

A reflection pulse is used to reverse the direction of propagation of the two wave packets simultaneously. The pulse is chosen to be 150 $\mu$s in duration with power 6.2 $\mu$W. This relatively long pulse reverses the momenta of the wave packets through a Bragg scattering process [17]. In response to the reflection pulse, the clouds turn around and propagate back toward their origin.

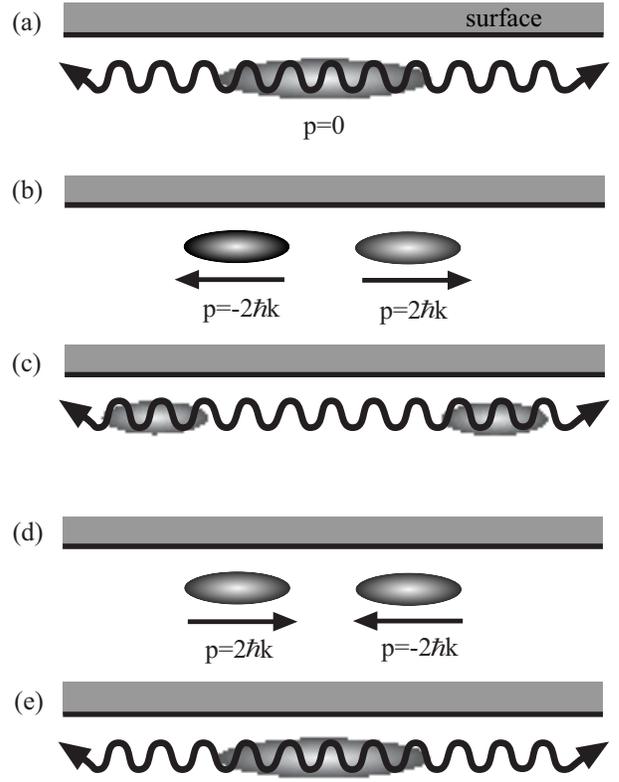

FIG. 2. (a) The beamsplitter pulse is turned on at t=0. (b) The condensate is split into two wave packets that propagate in the opposite direction at 0<t<T/2. (c) The reflection pulse is turned on at t=T/2. (d) The wave packets propagate back toward to the center of the waveguide at T/2<t<T. (e) The recombining pulse is turned on at t=T when the clouds are overlapped.

Upon the return to their origin, a second pair of splitting pulses serves to recombine the two condensate clouds. The schematic drawings in Fig. 2 show the whole sequence of splitting, reflecting, and recombining the condensate atoms by using standing-wave light fields. Like a simple optical beamsplitter/recombiner, our atom recombiner has two output ports. The first output port is represented by atoms having zero momentum while the second port is represented by atoms having $|p|=2\hbar k$. Since the splitting and recombining occur at the same spatial location, our atom interferometer is analogous to an optical Michelson interferometer. The relative phase shift between the two counterpropagating wave packets will change the fractional number of atoms in the two output ports. If the potential is perfectly symmetric, the relative phase shift should be zero, and the atoms should all be in the zero momentum state, as shown in Fig. 3(a). On the other hand, if the waveguide potential is not symmetric, the nonzero relative phase shift should lead to the presence of atoms in the $|p=\pm 2\hbar k\rangle$ state. In a special case where the relative phase shift is $\pi$, the number of atoms in the $|p=0\rangle$ state should be nearly zero due to destructive interference, and all the atoms will appear in $|p=\pm 2\hbar k\rangle$ state, as shown in Fig. 3(b). The populations in the two momentum states are anticorrelated since the total number is conserved.

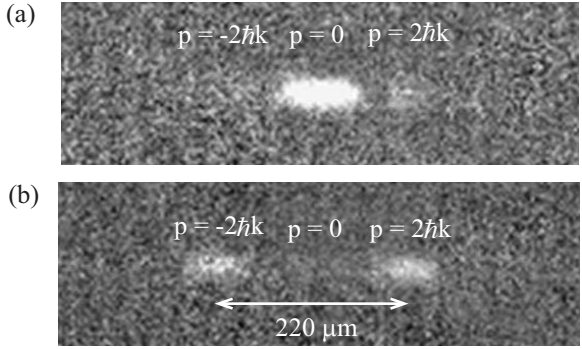

FIG. 3. Interference pattern of (a) phase shift = $2n\pi$ and (b) phase shift = $(2n+1)\pi$. The absorption images are taken 10 ms after the recombining pulse.

To demonstrate interference, a differential phase shift is introduced by a magnetic gradient while the two wave packets are spatially separated after the splitting pulse. The differential phase shift $\Delta\phi_{mag}$ can be written as

$$\Delta\phi_{mag} = \int_{t1}^{t2} \frac{\Delta H}{\hbar} dt = \int_{t1}^{t2} \frac{\mu_{Rb} B'(t)\Delta x(t)}{\hbar} dt ,$$

where $\mu_{Rb}$ is the magnetic dipole moment of $^{87}$Rb, $B'$ is a time-dependent magnetic gradient, $\Delta x$ is the time-dependent separation of the two clouds, $t_1$ is the time at which the magnetic gradient is turned on, and $t_2$ is the time at which the magnetic gradient is turned off. The magnetic gradient is provided by a single wire that is perpendicular to the waveguide and 3.6 mm away from its center. The interference due to the change of the magnetic gradient is shown in Fig. 4(a). The total propagation time of condensates in the waveguide is 1 ms, and the time between each pulse is 0.5 ms. The magnetic gradient is switched on 0.25 ms before the reflection pulse for a duration of 0.5 ms. The maximum separation of the two clouds is around 12 $\mu$m, which is small compared to the full width at half maximum of the cloud size, ~100 $\mu$m. The periodicity of the interference can be changed by varying the timing of the magnetic gradient pulse, which changes the separation of the clouds when the magnetic gradient is turned on. The result of the interference with the magnetic gradient switched on 0.05 ms before the reflection pulse is also shown in Fig. 4(b). The maximum contrast ratio shown in the figure is as large as 100%.

Interference is also observed after 10 ms propagation in the waveguide. An initial velocity of the condensate is created in the trap with a longitudinal frequency of 5 Hz, and the differential phase shift is observed by changing the total propagation time in the waveguide. The result of the interference is shown in Fig 5. The best contrast ratio of the interference at 10 ms, with maximum separation of the two clouds of about 120 $\mu$m, is 20%. The contrast ratio drops rapidly as we increase the propagation time in the waveguide. We attribute this reduction of the contrast to the dispersion of the wave packets. The dispersion might arise from atom-atom interaction, an inhomogeneous guiding potential, or a nonzero curvature of the applied magnetic gradient; this topic requires further study.

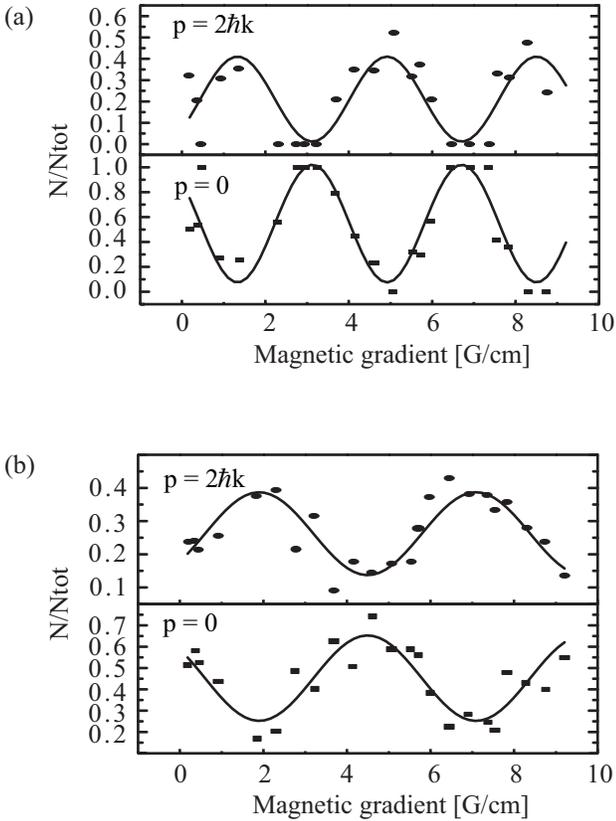

FIG. 4. Interference fringes after 1 ms propagation time in the waveguide with (a) the magnetic gradient turned for 500 $\mu$s while the average separation of clouds is 8.82 $\mu$m and with (b) the magnetic gradient turned for the same time while the average separation of clouds is 6.94 $\mu$m.

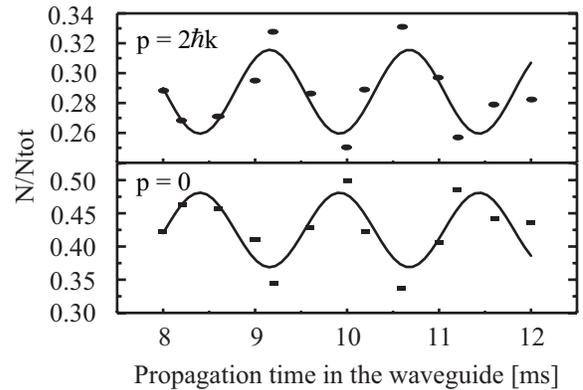

FIG. 5. Interference fringes after about 10 ms propagation in the waveguide.

In conclusion, we have demonstrated an on-chip atom Michelson interferometer. This is the first on-chip observation of atom interference between external states. The coherence of the wave packets has been observed up to 10 ms in the waveguide with a corresponding contrast ratio of 20%. The contrast ratio needs to be improved for a future interferometer design in which the cloud propagates for longer distances. The optical technique for splitting, reflecting, and recombining the condensates enables us to study decoherence effects that may affect coherent atom-chip devices in general.

The authors would like to thank P. D. D. Schwindt and L. Czaia for experimental assistance and useful discussion. This work was supported by the Army Research Office and the Office of the Secretary of Defense through a MURI program in Ultracold Atom Optics Science and Technology (Grant No. DAAD19-00-1-0163), the Defense Advanced Research Projects Agency's Defense Science Office through a PINS program (Grant No. W911NF-04-1-0043), the Office of Naval Research (Grant No. N00014-03-1-0551) and the National Science Foundation (Grant No. PHY-0096822).

---

*Present address: PRESTO, Japan Science and Technology Agency(JST), and Department of Applied Physics, School of Engineering, The University of Tokyo, Hongo 7-3-1, Bunkyo-ku, Tokyo 113-8656, JAPAN